\documentclass[12pt]{article}
\usepackage{a4wide}
\usepackage{axodraw}
\usepackage{latexsym}
\usepackage{epsfig}
\def\l{\langle}
\def\r{\rangle} 
\def\g{\gamma}

\begin{document}

\thispagestyle{empty}

\begin{flushright}
  NIKHEF/2000-005\\
  hep-ph/0003318
\end{flushright}

\vspace{1.5cm}

\begin{center}
  {\Large \bf Helicity Amplitudes for Single-Top Production}\\[.3cm]
  \vspace{1.7cm}
  {\sc J. van der Heide, E. Laenen, L. Phaf and  S. Weinzierl}\\
  \vspace{1cm}
  {\it NIKHEF Theory Group\\
    Kruislaan 409, 1098 SJ Amsterdam, The Netherlands} \\
\end{center}

\vspace{2cm}

\begin{abstract}\noindent
  {Single top quark production at hadron colliders allows a direct
    measurement of the top quark charged current coupling. We present
    the complete tree-level helicity amplitudes for four processes
    involving the production and semileptonic decay of a single top
    quark: $W$-gluon fusion, flavor excitation, $s$-channel production
    and $W$-associated production.  For the first three processes we
    study the quality of the narrow top width approximation.  We also
    examine momentum and angular distributions of some of the final
    state particles. }
\end{abstract}

\vspace*{\fill}


\newpage

\reversemarginpar

\section{Introduction}

Of the properties of the top quark, discovered at the Fermilab
Tevatron \cite{Abachi:1995iq, Abe:1995hr}, so far only its mass has 
been directly measured, and, to some extent, its QCD coupling
strength via the total production cross section.  There is hope to measure its
charged-current coupling, both in strength and handedness, via
single-top production at hadron colliders 
\cite{Dicus:1986wx} - \cite{Tait:1999cf}
in Run II at the Tevatron.

Besides having obvious intrinsic value, this measurement has
additional importance because the charged-current top coupling might
be particularly sensitive to certain signals of new physics
\cite{Carlson:1995ck}, \cite{Tait:1997fe}, \cite{Carlson:1994bg} - \cite{Lu:1997aa}.
Our context here is however the Standard Model.

To be sure, isolating a clear signal for single top production will not be
easy, in view of the complicated final state and the many backgrounds,
such as $t\bar{t}$ production. Valuable work has recently been done 
on constructing optimal signal
definitions, and examining corresponding backgrounds and acceptances
\cite{Stelzer:1998ni,Belyaev:1998dn,Liu:1999ra}.

There are various partonic subprocesses that lead to the
production of a single top. The ones for which we derive tree-level helicity 
amplitudes are the ``W-gluon fusion'' process
\begin{eqnarray}
\label{production1}
u + g \rightarrow t + d + \bar{b},
\end{eqnarray}
the ``flavor excitation'' process
\begin{eqnarray}
\label{production2}
u + b \rightarrow t + d,
\end{eqnarray}
the Drell-Yan like ``s-channel'' process, which occurs via a
virtual timelike W-boson
\begin{eqnarray}
\label{production3}
u + \bar{d} \rightarrow t + \bar{b},
\end{eqnarray}
and the ``W-associated'' production process
\begin{eqnarray}
  \label{production4}
g +  b \rightarrow t + W
\end{eqnarray}
with the W decaying hadronically. 
In reactions (\ref{production1}) and (\ref{production2}) it is
understood that we may replace the $(u,d)$-quark pair by
$(\bar{d},\bar{u})$, $(c,s)$ and $(\bar{s},\bar{c})$. In reaction
(\ref{production3}) we may replace the $(u,\bar{d})$-pair by
$(c,\bar{s})$. In addition, CKM suppressed combinations 
may be included.

Because of its large mass, the electroweak decay of the top quark
proceeds so rapidly that top bound states do not have time to
form \cite{Bigi:1986jk}. This also means that the decay products of the top quark are
correlated with its spin. It is therefore desirable to include the
semileptonic decay\footnote{Single-top production with hadronic top
decay suffers from large QCD backgrounds.}
of the top quark in the amplitudes:
\begin{eqnarray}
\label{decay}
t \rightarrow b + \bar{l} + \nu.
\end{eqnarray}

The complete tree-level amplitudes for flavor-excitation and 
$s$-channel production can be obtained by crossing from those for
\begin{eqnarray}
\label{amp1}
0 \rightarrow b + \bar{l} + \nu + \bar{b} + \bar{u} + d.
\end{eqnarray}
with ``$0$'' representing the vacuum.
Likewise, the amplitudes for W-gluon fusion and W-associated production are obtained 
by crossing from
\begin{eqnarray}
\label{amp2}
0 \rightarrow b + \bar{l} + \nu + \bar{b} + \bar{u} + d + g
\end{eqnarray}

In this paper we present the complete tree-level helicity amplitudes
for (\ref{amp1}) and (\ref{amp2}). We obtain compact expressions by using
spinor helicity methods. 
We present in addition all amplitudes for the subprocesses
(\ref{production1}-\ref{production4}) in the narrow top width
approximation \cite{Pittau:1996rp,Mahlon:1995us}.  This
allows us to check the quality of this approximation
against the full calculation, for which only a few of the diagrams actually
involve a top quark. 

Although the helicity amplitudes for 
(\ref{production1}-\ref{production4}), including top decay,
may be obtained as well as Fortran code from the program Madgraph \cite{Stelzer:1994ta},
we believe that our analytical results are valuable for a number of
reasons. First, analytical expressions can offer additional
insights, e.g. in the phase space structure of the cross section
near the top mass pole.
Second, they lead to even more compact computer codes by 
allowing numerical crossing, and 
third, they allow evaluation in terms of spinor products,
and in different computer languages (we use C++). Finally, our
results constitute a nontrivial check on this useful program.
In this context we also mention the program Onetop \cite{Carlson:1995cm}, 
and the general purpose programs Pythia
\cite{Sjostrand:1994yb} and Herwig \cite{Marchesini:1992ch}
which can be used as event generators for single top production.

In order to test these amplitudes numerically, we study
various distributions in momentum and angle
of some of the final state particles for each
process separately. We limit ourselves here to the subprocesses 
(\ref{production1}-\ref{production3}), because process (\ref{production4}) 
is negligible \cite{Heinson:1997zm,Stelzer:1998ni}
at the Tevatron. We note that whether a subprocess is a 
leading order contribution or a higher order 
correction to another, depends on the definition of the final state.  
Thus, for sufficiently inclusive quantities the W-gluon fusion
and flavor excitation processes are not independent:
a part of the former is then in fact a higher order QCD correction to the latter, and must
be mass factorized \cite{Dicus:1986wx}.
A similar argument applies to W-gluon fusion
and the $s$-channel process in the case where the gluon couples to the 
$u$-$d$-quark line, and the signal is defined to be inclusive
with respect to the presence of the light quark jet. Furthermore, whether
bottom quarks are part of the initial state is a choice of
factorization scheme. We adopt a five-flavor scheme and
include bottom quark parton distribution functions.
In this paper we wish to examine some characteristics of each process individually,
rather than perform a comprehensive phenomenological study involving combinations
of these processes and their backgrounds, as such studies already exists in the literature
\cite{Stelzer:1998ni,Belyaev:1998dn}. 
We therefore focus on exclusive quantities, e.g. we require exactly three
jets or exactly two jets.

Our conventions for spinor helicity methods are listed
in section 2. Sections 3 and 4, together with appendices A and B, 
contain the helicity amplitudes for processes (\ref{amp1}) and (\ref{amp2}).
In section 5 we discuss the narrow top width approximation,
while the results of our  numerical studies can be found
in section 6. We conclude in section 7.

\section{Spinor helicity}

To compute the amplitudes for (\ref{amp1}) and (\ref{amp2}) 
we use spinor helicity methods 
\cite{Berends:1981rb} - \cite{Gastmans:1990xh}.
We limit ourselves here to listing our conventions,
for reviews of spinor helicity methods see e.g.
\cite{Mangano:1990by,Dixon:1996wi}. 

With spinor helicity methods we can express scattering
amplitudes in terms of massless Weyl spinors 
of helicity $\pm\frac{1}{2}$ 
\begin{eqnarray}
  \label{eq:14}
u(p,\pm) = v(p,\mp) =  | p \pm \r \,, \qquad   \bar{u}(p,\pm) = \bar{v}(p,\mp) = \l p \pm |  \,.
\end{eqnarray}
External fermion states are directly expressed in terms of these.
Our convention is to take all particles outgoing. 
For example an outgoing massless fermion with positive helicity is denoted by $\l p + |$, while an
outgoing massless antifermion with positive helicity is denoted by $| p - \r$.
The gluon polarization vectors, of helicity $\pm 1$, may 
be written as
\begin{eqnarray} \label{eq:1}
\varepsilon^+_\mu(k,q) = \frac{\l q- | \g_\mu | k- \r}{\sqrt{2}\l qk \r}, & &
\varepsilon^-_\mu(k,q) = \frac{\l q+ | \g_\mu | k+ \r}{\sqrt{2} [ kq ]}.
\end{eqnarray}
We have used the customary short-hand notation:
\begin{eqnarray}\label{eq:2}
\l ij \r = \l p_i - | p_j + \r, & & [ ij ] = \l p_i + | p_j - \r .
\end{eqnarray}
In (\ref{eq:1}) $k$ is the gluon momentum and $q$  an arbitrary light-like
``reference momentum''.  The dependence on the choice of $q$ drops out
in gauge-invariant amplitudes.  We shall also employ the abbreviations
\begin{eqnarray}\label{eq:3}
\l i- | k+l | j- \r & = & \l ik \r [ kj] + \l il \r [ lj ], \nonumber \\
s_{ij...k} & = & \left( p_i + p_j + ... + p_k \right)^2  ,
\end{eqnarray}
with all momenta null-vectors.

To investigate the narrow-width approximation (section 5),
we must treat the massive top quark as an
external state. Appropiately extended spinor techniques exist
(\cite{Kleiss:1985yh}, \cite{Berends:1985gf} - \cite{Dittmaier:1998nn}).  
Even though helicity is not a conserved quantum number for a massive
particle, a massive positive-energy spinor satisfying the Dirac equation has a two-fold
degeneracy (e.g. labelled by a spin-component quantized along some axis). 
With slight abuse of notation we label these two states  by
``$+$'' and ``$-$''.  Let $p$ be a four-vector with $p^2=m^2$ and
$p_0 > 0$, and let $q$ be an arbitrary null vector with $q_0 > 0$.  We define
\begin{eqnarray}\label{eq:4}
u(p,+) = \frac{1}{\sqrt{2 p q}} \left( p\!\!\!/ + m \right) | q - \r,
& & 
v(p,+) = \frac{1}{\sqrt{2 p q}} \left( p\!\!\!/ - m \right) | q - \r, \nonumber \\
u(p,-) = \frac{1}{\sqrt{2 p q}} \left( p\!\!\!/ + m \right) | q + \r,
& & 
v(p,-) = \frac{1}{\sqrt{2 p q}} \left( p\!\!\!/ - m \right) | q + \r.
\end{eqnarray}
For the conjugate spinors we have
\begin{eqnarray}\label{eq:5}
\bar{u}(p,+) = \frac{1}{\sqrt{2 p q}} \l q - | \left( p\!\!\!/ + m \right), 
& & 
\bar{v}(p,+) = \frac{1}{\sqrt{2 p q}} \l q - | \left( p\!\!\!/ - m \right), \nonumber \\
\bar{u}(p,-) = \frac{1}{\sqrt{2 p q}} \l q + | \left( p\!\!\!/ + m \right),
& & 
\bar{v}(p,-) = \frac{1}{\sqrt{2 p q}} \l q + | \left( p\!\!\!/ - m \right). 
\end{eqnarray}
It is easy to check that for these spinors the Dirac equations, 
orthogonality and completeness relations hold.
A massive quark propagator may be expressed in terms of 
massless Weyl spinors via
\begin{eqnarray}\label{eq:74}
\frac{i \delta_{ij}}{p^2-m^2+i\epsilon} \left( p_+ + p_- + m \right),
\end{eqnarray}
where $p_+$ and $p_-$ are of the form
\begin{eqnarray}\label{eq:75}
p_+ & = & | p_1+ \r \l p_1\!+\! | \;+ ... +\; | p_n+ \r \l p_n\!+ \!| \nonumber \\
p_- & = & | p_1- \r \l p_1\!-\! | \;+ ... +\; | p_n- \r \l p_n\!- \!| 
\end{eqnarray}
for some nullvectors $p_1$,...,$p_n$.

For all other vertices and propagators we use 
standard Feynman rules, in the conventions of \cite{Bohm:1986rj},
and the 't Hooft-Feynman $R_\xi$-gauge with $\xi=1$.
We neglected all fermion masses except
the top mass. As a consequence, neither diagrams containing a Higgs boson 
nor diagrams with would-be Goldstone bosons contribute.

\section{W-gluon fusion and W-associated production}

In this section we present the helicity amplitudes for the process
\begin{eqnarray}
\label{reaction3}
0 \rightarrow \nu(p_1) + \bar{l}(p_2) + b(p_3) + \bar{b}(p_4) + g(p_5) + d(p_6) + \bar{u}(p_7).
\end{eqnarray}
The amplitudes are calculated in tree approximation 
at order $O(g g_w^4)$, where $g$ denotes the strong coupling and $g_w$
the electroweak coupling.  There are also tree
diagrams of order $O(g^3 g_w^2)$ (``QCD + Weak'') contributing to
(\ref{reaction3}). This gauge-invariant set of graphs does not contain a 
top quark as an intermediate state, and we do not consider it in this paper. 

The results for the (W-gluon fusion) processes 
$u+g \rightarrow \nu +\bar{l} + b + \bar{b} + d $ 
and 
$\bar{d} + g \rightarrow \nu +\bar{l} + b + \bar{b} + \bar{u}$, 
as well as for the (W-associated) process
$b +g \rightarrow \nu +\bar{l} + b + d + \bar{u} $,
can be obtained from those of process
(\ref{reaction3}) by crossing, under which
the crossed momentum and helicity change sign.
The color decomposition for the amplitude 
(\ref{reaction3}) reads
\begin{eqnarray} \label{eq:9}
A_{Wg} & = & g T^a_{34} \delta_{67} A^{(1)}_{Wg} + g \delta_{34} T^a_{67} A^{(2)}_{Wg}.
\end{eqnarray}
The partial amplitudes $A^{(1)}_{Wg}$ and $A^{(2)}_{Wg}$ are
gauge-invariant by themselves.  $A^{(1)}_{Wg}$ corresponds to
diagrams where the gluon couples to the $b$-$\bar{b}$-fermion line,
whereas $A^{(2)}_{Wg}$ involves the gluon coupling to the
$d$-$\bar{u}$-fermion line.  Representative Feynman diagrams for the
partial amplitudes $A^{(1)}_{Wg}$ and $A^{(2)}_{Wg}$ are shown in 
Fig.~\ref{figure_a} and Fig.~\ref{figure_b}, respectively. There are 21
diagrams contributing to $A^{(1)}_{Wg}$, 3 of them contain a
top quark. The partial amplitude $A^{(2)}_{Wg}$ is made up of 24
diagrams, with 2 containing a top quark.  

For the color matrices we have used the short-hand notation $T^a_{34} = T^a_{i_3  j_4}$.
They are normalized as
\begin{eqnarray} \label{eq:10}
\mbox{Tr}\; T^a T^b & = & \frac{1}{2} \delta^{ab}.
\end{eqnarray}
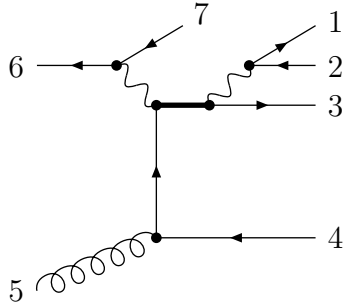
\begin{figure}
\begin{center}
\begin{picture}(100,100)(0,0)
\ArrowLine(90,20)(30,20)
\ArrowLine(30,20)(30,70)
\ArrowLine(50,70)(90,70)
\SetWidth{2}
\Line(30,70)(50,70)
\SetWidth{0.5}
\Vertex(30,20){2}
\Vertex(30,70){2}
\Vertex(50,70){2}
\Vertex(15,85){2}
\Vertex(65,85){2}
\ArrowLine(40,100)(15,85)
\ArrowLine(15,85)(-15,85)
\Photon(15,85)(30,70){3}{2}
\Photon(50,70)(65,85){3}{2}
\Gluon(-15,0)(30,20){4}{5}
\ArrowLine(90,85)(65,85)
\ArrowLine(65,85)(90,100)
\Text(95,102)[l]{$1$}
\Text(95,85)[l]{$2$}
\Text(95,70)[l]{$3$}
\Text(95,20)[l]{$4$}
\Text(-20,0)[r]{$5$}
\Text(-20,85)[r]{$6$}
\Text(50,105)[r]{$7$}
\end{picture}
\caption{\label{figure_a} \it A representative Feynman diagram for single-top production contributing to the
  partial amplitude $A^{(1)}_{Wg}$ (\ref{eq:9}). The top quark line is thickened.}
\end{center}
\end{figure}
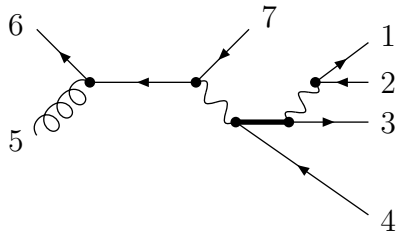
\begin{figure}
\begin{center}
\begin{picture}(100,100)(0,0)
\ArrowLine(80,100)(60,80)
\ArrowLine(60,80)(20,80)
\ArrowLine(20,80)(0,100)
\Gluon(0,60)(20,80){4}{3}
\Photon(60,80)(75,65){3}{2}
\ArrowLine(125,30)(75,65)
\ArrowLine(95,65)(125,65)
\SetWidth{2}
\Line(75,65)(95,65)
\SetWidth{0.5}
\Photon(95,65)(105,80){3}{2}
\ArrowLine(125,80)(105,80)
\ArrowLine(105,80)(125,95)
\Vertex(60,80){2}
\Vertex(20,80){2}
\Vertex(75,65){2}
\Vertex(95,65){2}
\Vertex(105,80){2}
\Text(-5,102)[r]{$6$}
\Text(-5,58)[r]{$5$}
\Text(85,105)[l]{$7$}
\Text(130,28)[l]{$4$}
\Text(130,65)[l]{$3$}
\Text(130,80)[l]{$2$}
\Text(130,98)[l]{$1$}
\end{picture}
\caption{\label{figure_b} \it A representative Feynman diagram for single-top production contributing to the 
  partial amplitude $A^{(2)}_{Wg}$ (\ref{eq:9}). The top quark line is thickened.}
\end{center}
\end{figure}
We can decompose the two subamplitudes in (\ref{eq:9}) further according
to their electroweak structure:
\begin{eqnarray} \label{eq:11}
A^{(1)}_{Wg} & = & \frac{e^4 V_{ud}^\ast}{2 \sin^2 \theta_W}
\left( \frac{||V_{tb}||^2}{2 \sin^2 \theta_W} A^{(1,1)}_{Wg}
      + \left( v^\gamma_d v^\gamma_b + v^Z_d v^Z_b {\cal P}_Z(s_{345}) \right) A^{(1,2)}_{Wg}
\right. \nonumber \\
& & \left.      + \left( v^\gamma_u v^\gamma_b + v^Z_u v^Z_b {\cal P}_Z(s_{345}) \right) A^{(1,3)}_{Wg}
      + \left( v^\gamma_b - \frac{\cos \theta_W}{\sin \theta_W} v^Z_b {\cal P}_Z(s_{345}) \right) A^{(1,4)}_{Wg}
\right. \nonumber \\
& & \left. + \left( v_e^\gamma v_b^\gamma + v_e^Z v_b^Z {\cal P}_Z(s_{345}) \right) A^{(1,5)}_{Wg}
+ v_\nu^Z v_b^Z {\cal P}_Z(s_{345}) A^{(1,6)}_{Wg}
\right), \nonumber \\
A^{(2)}_{Wg} & = & \frac{e^4 V_{ud}^\ast}{2 \sin^2 \theta_W}
\left( \frac{||V_{tb}||^2}{2 \sin^2 \theta_W} A^{(2,1)}_{Wg}
      + \left( v^\gamma_d v^\gamma_b + v^Z_d v^Z_b {\cal P}_Z(s_{34}) \right) A^{(2,2)}_{Wg}
\right. \nonumber \\
& & \left.      + \left( v^\gamma_u v^\gamma_b + v^Z_u v^Z_b {\cal P}_Z(s_{34}) \right) A^{(2,3)}_{Wg}
      + \left( v^\gamma_b - \frac{\cos \theta_W}{\sin \theta_W} v^Z_b {\cal P}_Z(s_{34}) \right) A^{(2,4)}_{Wg}
\right. \nonumber \\
& & \left. + \left( v_e^\gamma v_b^\gamma + v_e^Z v_b^Z {\cal P}_Z(s_{34}) \right) A^{(2,5)}_{Wg}
+ v_\nu^Z v_b^Z {\cal P}_Z(s_{34}) A^{(2,6)}_{Wg}
\right).
\end{eqnarray}
Here
\begin{eqnarray} \label{eq:12}
{\cal P}_Z(s)  =  \frac{s}{s - m_Z^2 }, & & 
e = g_w \sin \theta_W \nonumber \\
v^{\gamma,L}_f = - Q, & & v^{\gamma,R}_f = - Q \nonumber\\
v^{Z,L}_f = \frac{I_3 - Q \sin^2 \theta_W}{\sin \theta_W \cos \theta_W},
& & 
v^{Z,R}_f = \frac{- Q \sin \theta_W}{\cos \theta_W},
\end{eqnarray}
where $Q$ and $I_3$ denote the charge and the third component of the weak
isospin of the fermion.
The labels $L$ and $R$ denote the left- and right-handed couplings. 
Further, $\theta_W$ denotes the Weinberg angle and $V_{ud}$ and $V_{tb}$ denote CKM-matrix elements.\\
\\
Because the W-boson only couples to left-handed fermions, all non-vanishing
amplitudes have the helicity configuration $(p_1^-, p_2^+, p_6^-,
p_7^+)$.  Furthermore the helicity along the fermion line
$b(p_3)$-$\bar{b}(p_4)$ is conserved.
Due to their number and length we have collected the explicit expressions for
the partial helicity amplitudes $A_{Wg}^{(k,l)}$ in (\ref{eq:11})
in appendix A.  We have verified 
the correctness of these expressions
by numerical comparison with the computer code generated by Madgraph \cite{Stelzer:1994ta}.

The cross section for W-gluon fusion, summed and averaged over helicities and colours
is then given by
\begin{eqnarray}
\sigma_{Wg} & = & \sum\limits_{i,j} \int dx_1 f_i(x_1) \int dx_2 f_j(x_2) \int d\phi_5
\frac{1}{8\hat{s}} \sum\limits_{helicities} \frac{1}{2} \left(
 \left|A^{(1)}_{Wg} \right|^2 + \left|A^{(2)}_{Wg} \right|^2 \right) \Theta(\mbox{cuts}) \nonumber \\
\end{eqnarray}  
where $f_i(x_1)$ and $f_j(x_2)$ are the parton densities of the initial partons $i$ and $j$,
$d\phi_5$ denotes the phase space measure for five massless particles, $2\hat{s}$ is the flux factor,
$\Theta(\mbox{cuts})$ represents the jet-defining cuts and  $A^{(1)}_{Wg}$ and $A^{(2)}_{Wg}$
are the amplitudes in (\ref{eq:11}) with partons $i$ and $j$ crossed into the initial state.

\section{Flavor excitation and $s$-channel}

The helicity amplitudes for both these processes can be obtained by crossing from
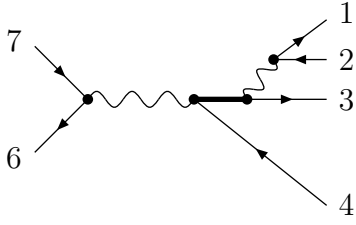
\begin{figure}
\begin{center}
\begin{picture}(100,100)(0,-20)
\ArrowLine(0,40)(20,20)
\ArrowLine(20,20)(0,0)
\Photon(20,20)(60,20){3}{3}
\ArrowLine(110,-20)(60,20)
\ArrowLine(80,20)(110,20)
\SetWidth{2}
\Line(60,20)(80,20)
\SetWidth{0.5}
\Photon(80,20)(90,35){3}{2}
\ArrowLine(110,35)(90,35)
\ArrowLine(90,35)(110,50)
\Vertex(20,20){2}
\Vertex(60,20){2}
\Vertex(80,20){2}
\Vertex(90,35){2}
\Text(-5,42)[r]{$7$}
\Text(-5,-2)[r]{$6$}
\Text(115,-20)[l]{$4$}
\Text(115,20)[l]{$3$}
\Text(115,35)[l]{$2$}
\Text(115,53)[l]{$1$}
\end{picture}
\caption{\label{figure_c} \it A representative Feynman diagram corresponding to
the partial amplitude $A^{(1)}_{Wb}$ (\ref{amp_Wb}).
The top quark line is thickened.}
\end{center}
\end{figure}
\begin{eqnarray} 
\label{reaction5}
0 \rightarrow \nu(p_1) + \bar{l}(p_2) + b(p_3) + \bar{b}(p_4) + d(p_6) + \bar{u}(p_7).
\end{eqnarray}
The tree-level amplitudes  correspond
to order $O(g_w^4)$. To this order the color decomposition
of the amplitude for (\ref{reaction5}) is simply
\begin{eqnarray} 
\label{amp_Wb}
A_{Wb} & = & \delta_{34} \delta_{67} A^{(1)}_{Wb}
\end{eqnarray}
since no gluons are involved.  A representative Feynman diagram is
shown in Fig.~\ref{figure_c}.  There are 10 Feynman diagrams
contributing to this amplitude, only one of them contains a top quark.  We
write the partial amplitudes $A^{(1)}_{Wb}$ as
\begin{eqnarray} \label{eq:16}
A^{(1)}_{Wb} & = & \frac{e^4 V_{ud}^\ast}{2 \sin^2 \theta_W}
\Biggl( \frac{||V_{tb}||^2}{2 \sin^2 \theta_W} A^{(1,1)}_{Wb}
      + \left( v^\gamma_d v^\gamma_b + v^Z_d v^Z_b {\cal P}_Z(s_{34}) \right) A^{(1,2)}_{Wb}
 \nonumber \\
& &       + \left( v^\gamma_u v^\gamma_b + v^Z_u v^Z_b {\cal P}_Z(s_{34}) \right) A^{(1,3)}_{Wb}
      + \left( v^\gamma_b - \frac{\cos \theta_W}{\sin \theta_W} v^Z_b {\cal P}_Z(s_{34}) \right) A^{(1,4)}_{Wb}
 \nonumber \\
& &  + \left( v_e^\gamma v_b^\gamma + v_e^Z v_b^Z {\cal P}_Z(s_{34}) \right) A^{(1,5)}_{Wb}
+ v_\nu^Z v_b^Z {\cal P}_Z(s_{34}) A^{(1,6)}_{Wb}
\Biggr) \,.
\end{eqnarray}
All helicity amlitudes have the helicities $(p_1^-, p_2^+)$ and
$(p_6^-, p_7^+)$. Again the helicity along the
$b$-$\bar{b}$-line is conserved.  The explicit expressions for the
$A^{(1,k)}_{Wb}$ in (\ref{eq:16}) are collected in appendix B.  

We have checked these results numerically 
with the computer code produced by Madgraph \cite{Stelzer:1994ta},
and found agreement. For the present process we could in addition
compare to results produced by the Comphep program \cite{Pukhov:1999gg}, and
found agreement as well.
With the helicity amplitudes at hand one obtains the cross section for 
flavor excitation, summed and averaged over helicities and colors,
as
\begin{eqnarray}
\sigma_{Wb} & = & \sum\limits_{i,j} \int dx_1 f_i(x_1) \int dx_2 f_j(x_2) \int d\phi_4
\frac{1}{8\hat{s}} \sum\limits_{helicities} 
 \left|A^{(1)}_{Wb} \right|^2 \Theta(\mbox{cuts}) 
\end{eqnarray}  
with $d\phi_4$ the phase space measure for 4 massless particles and $i$ and $j$ label the partons crossed
for flavor excitation. The cross section for
the s-channel process is obtained in a similar way by crossing the appropriate partons.

There are tree diagrams of order $O(g^2 g_w^2)$ 
that contribute to (\ref{reaction5}). As they are a separately
gauge-invariant set and do not contain a top quark they are not
directly relevant to us here.
However, for this case we do present these ``QCD + Weak''-amplitudes 
because they may be easily obtained from the $O(g_w^4)$ tree amplitudes, as follows
\begin{eqnarray} \label{eq:17}
A_{\rm QCD+Weak} & = & \frac{1}{2} \left( \delta_{37} \delta_{64} -\frac{1}{N_c} \delta_{34} \delta_{67} \right)
\frac{g^2 e^2 V_{ud}^\ast}{2 \sin^2 \theta_W} \left( A^{(1,2)}_{Wb} + A^{(1,3)}_{Wb} \right),
\end{eqnarray}
where $N_c$ denotes the number of colors. In contrast to the process of the
previous section, these ``QCD + Weak''-amplitudes do not interfere with
the $O(g_w^4)$ amplitudes.

\section{Narrow width approximation}

By including the top quark semileptonic decay in the amplitudes for
(\ref{amp1}) and (\ref{amp2}), we must include as well many diagrams in
which no top is present. Therefore it is interesting to know
to what extent results are approximated by producing the top quark as
an on-shell particle, whose decay happens independently from its
production.

A numerical indication that the narrow top width approximation for the W-gluon,
s-channel and W-associated processes works well is already present in
\cite{Stelzer:1998ni}. Here we examine this issue both numerically
and analytically for the three subprocesses (\ref{production1}),
(\ref{production2})\footnote{For this channel the narrow width
approximation has even been examined to one loop in \cite{Pittau:1996rp}.}
and (\ref{production3}).
Thus we need the helicity amplitudes for the W-gluon fusion process without top decay
\begin{eqnarray} 
\label{reaction4}
0 \rightarrow t(p_8) + \bar{b}(p_4) + g(p_5) + d(p_6) + \bar{u}(p_7).
\end{eqnarray}
The amplitude may again be color-decomposed as
\begin{eqnarray} \label{eq:19}
A_{Wg,prod} & = & g T^a_{84} \delta_{67} A^{(1)}_{Wg,prod} + g \delta_{84} T^a_{67} A^{(2)}_{Wg,prod}
\end{eqnarray}
with
\begin{eqnarray} \label{eq:20}
A^{(1)}_{Wg,prod} & = & \frac{e^2 V_{ud}^\ast V_{tb}}{2 \sin^2 \theta_W} \cdot
\frac{(-i) 2 \sqrt{2}}{s_{67} -m_W^2}
\frac{B^{(1)}_{Wg,prod}}{\sqrt{ - \l 2- | 4+5+6+7 |2- \r}}, \nonumber \\
A^{(2)}_{Wg,prod} & = & \frac{e^2 V_{ud}^\ast V_{tb}}{2 \sin^2 \theta_W} \cdot
\frac{(-i) 2 \sqrt{2}}{s_{567} -m_W^2}
\frac{B^{(2)}_{Wg,prod}}{\sqrt{ - \l 2- | 4+5+6+7 |2- \r}}.
\end{eqnarray}
As reference momentum for the massive spinor we have chosen $q=p_2$.
As before all non-vanishing amplitudes have the helicity configuration
$(p_6^-,p_7^+)$.  The non-vanishing amplitudes are
\begin{eqnarray} \label{eq:21}
B^{(1)}_{Wg,prod}(p_4^+,p_5^+,p_8^-) & = & 
\frac{\l 6- | 4+5+7 | 2- \r}{\l 65 \r} 
\left( \frac{\l 6- | 4+5 | 7- \r}{\l 45 \r} + \frac{[ 74 ] \l 6- | 4+7 | 5- \r}{s_{467} -m^2} \right),
\nonumber \\ 
B^{(1)}_{Wg,prod}(p_4^+,p_5^+,p_8^+) & = & 
- \frac{m \l 26 \r}{\l 65 \r}
\left( \frac{\l 6- | 4+5 | 7- \r}{\l 45 \r} + \frac{[ 74 ] \l 6- | 4+7 | 5- \r}{s_{467} -m^2} \right),
\nonumber \\ 
B^{(1)}_{Wg,prod}(p_4^+,p_5^-,p_8^-) & = &
\frac{[ 74 ]}{ [ 54 ] \left( s_{467} - m^2 \right)}
\left( \l 5- | 4+6+7 | 2- \r [ 47 ] \l 76 \r + m^2 [ 24 ] \l 56 \r \right),
\nonumber \\ 
B^{(1)}_{Wg,prod}(p_4^+,p_5^-,p_8^+) & = &
- \frac{m}{s_{467} -m^2} \frac{ [ 47 ] }{ [ 45 ] }
\left( \l 25 \r \l 67 \r [ 74 ] + \l 56 \r \l 2- | 5+6+7 | 4- \r \right),
\end{eqnarray}
\begin{eqnarray} \label{eq:22}
B^{(2)}_{Wg,prod}(p_4^+,p_5^+,p_8^-) & = & 
\frac{\l 6- | 4+5+7 | 2- \r \l 6- | 5+7 | 4- \r}{\l 56 \r \l 75 \r}, \nonumber \\
B^{(2)}_{Wg,prod}(p_4^+,p_5^+,p_8^+) & = & 
\frac{m \l 62 \r \l 6- | 5+7 | 4- \r}{\l 56 \r \l 75 \r}, \nonumber \\
B^{(2)}_{Wg,prod}(p_4^+,p_5^-,p_8^-) & = & 
\frac{[ 74 ] \l 2+ | (4+5+6+7) ( 5+6) | 7- \r}{ [ 57 ] [ 56 ]}, \nonumber \\
B^{(2)}_{Wg,prod}(p_4^+,p_5^-,p_8^+) & = & 
\frac{m [ 47 ] \l 2- | 5+6 | 7- \r}{ [ 57 ] [ 56 ]}.
\end{eqnarray}
The helicity amplitudes for
flavor-excitation and the s-channel process without the top decay
\begin{eqnarray} 
\label{reaction6}
0 \rightarrow t(p_8) + \bar{b}(p_4) + d(p_6) + \bar{u}(p_7),
\end{eqnarray}
are relatively simple and are given by
\begin{eqnarray} \label{eq:24}
A_{Wb,prod} & = & \delta_{84} \delta_{67} \frac{e^2 V_{ud}^\ast V_{tb}}{2 \sin^2 \theta_W} \cdot
\frac{2 i}{s_{67} -m_W^2}
\frac{B^{(1)}_{Wb,prod}}{\sqrt{ - \l 2- | 4+6+7 |2- \r}}.
\end{eqnarray}
As reference momentum for the massive spinor we have again chosen $q =
p_2$.  All helicity amplitudes have the configuration $(p_6^-,p_7^+)$.
The non-vanishing amplitudes are
\begin{eqnarray} \label{eq:25}
B^{(1)}_{Wb,prod}(p_4^+,p_8^-) & = & [ 47 ] \l 6- | 4+7 | 2- \r, \nonumber \\
B^{(1)}_{Wb,prod}(p_4^+,p_8^+) & = & m \l 26 \r [ 74 ].
\end{eqnarray}
Finally, let us give the amplitude for the top decay
\begin{eqnarray} \label{eq:26}
t(p_8) \rightarrow \nu(p_1) + \bar{l}(p_2) + b(p_3) .
\end{eqnarray}
With the choice $q=p_2$ as reference momentum for the top-spinor the
only non-vanishing amplitude is
\begin{eqnarray} \label{eq:27}
A_{dec}(p_1^-, p_2^+, p_3^-, p_8^-) & = & \frac{e^2 V_{tb}^\ast}{2 \sin^2
  \theta_W} \frac{2i}{s_{12} - m_W^2} \l 31 \r
\sqrt{ \l 2- | 1+3 | 2- \r}
\end{eqnarray}
To implement the narrow top width approximation we keep in the amplitude only terms
with a propagator $1/(p_8^2-m^2+ i m \Gamma)$ with 
$\Gamma$ the inclusive top width. For the amplitude squared we obtain
\begin{eqnarray} \label{eq:28}
| A |^2 & = & \left| \sum\limits_{\lambda} A_{dec}(...,p_8^\lambda) \frac{i}{p_8^2 - m^2 + i m \Gamma}
A_{prod}(...,p_8^\lambda) \right|^2 \nonumber \\
& = & | A_{dec}(...,p_8^-) |^2 \frac{1}{(p_8^2-m^2)^2+m^2 \Gamma^2} | A_{prod}(...,p_8^-)|^2,
\end{eqnarray}
because, with our choice of reference momentum for the massive spinor,
$A_{dec}(...,p_8^+)=0$.  In the limit of vanishing top width
the Breit-Wigner function in (\ref{eq:28}) reduces to a Dirac
delta-function and we obtain for the squared amplitude
\begin{eqnarray} \label{eq:29}
\frac{\pi}{m \Gamma} \delta(p_8^2 - m^2) | A_{dec}(...,p_8^-)|^2 | A_{prod}(...,p_8^-)|^2
\end{eqnarray}
The full $n$-particle phase space may be factorized accordingly
\begin{eqnarray} \label{eq:30}
d \phi_n(Q \rightarrow k_1,...,k_n) = \frac{1}{2\pi} d\phi_{n-2}(Q \rightarrow p_8, k_4, ... , k_n) 
dp_8^2 d\phi_{3}(p_8 \rightarrow k_1, k_2, k_3),
\end{eqnarray}
with $n=5$ for the W-gluon fusion and W-associated production,
and $n=4$ for flavor excitation and $s$-channel production.
Note that
\begin{eqnarray}
  \label{eq:76}
\int | A_{dec}(...,p_8^-)|^2 d\phi_{3}(p_8 \rightarrow k_1, k_2, k_3) = 2 m \Gamma_{\nu\bar{l}b}.
\end{eqnarray}
Numerical results for the narrow top-width approximation are presented in the next section.

\section{Numerical studies}

As announced in the introduction, we consider for the purposes
of numerical studies each subprocess separately.  
This is equivalent to assuming a hermetic detector with perfect
momentum resolution capable of distinguishing these three processes.  

Before describing our numerical studies we list our default
choices for physical constants and parameters.
For the masses and widths of the
electroweak bosons we use $m_Z = 91.187 \; \mbox{GeV}$, $\Gamma_Z = 2.49 \;
\mbox{GeV}$, $m_W = 80.41 \; \mbox{GeV}$ and $\Gamma_W = 2.06 \;
\mbox{GeV}$. For the top quark mass we use $m_{t} = 174 \;
\mbox{GeV}$. The width of the top quark is then calculated as
$\Gamma_t = 1.76 \; \mbox{GeV}$.  We use the leading order expression
for the running of the strong coupling constant:
\begin{eqnarray} \label{eq:31}
\alpha_s(\mu) & = & \alpha_s(m_Z) \left[1 + \frac{\alpha_s(m_Z)}{4 \pi} \beta_0 \ln \frac{\mu^2}{m_Z^2}\right]^{-1},
\end{eqnarray}
where $\beta_0 = 11 - \frac{2}{3} N_f$ and $N_f=5$.  We use the CTEQ4L
set for the parton densities \cite{Lai:1997mg} and we take therefore
$\alpha_s(m_Z) = 0.132$.  The running of the finestructure constant is
taken in account according to
\begin{eqnarray} \label{eq:32}
\alpha(\mu) & = & \alpha(0)\left[1-\Delta\alpha(m_Z)-\frac{\alpha(0)}{3\pi}\left(
\frac{20}{3}\ln \frac{\mu^2}{m_Z^2} - \frac{4}{15} \frac{(\mu^2-m_Z^2)}{m_t^2} \right)\right]^{-1},
\end{eqnarray}
with $\alpha(0)= 1/137.036$ and $\Delta\alpha(m_Z)=0.059363$
(\cite{Martin:1995we} - \cite{Alemany:1997tn}).
We consider $p$-$\bar{p}$ collisions with a
center-of-mass energy $\sqrt{S} = 2.0 \;\mbox{TeV}$ (Tevatron) and $p$-$p$
collisions with a center-of-mass energy $\sqrt{S} = 14 \;\mbox{TeV}$ (LHC).
For the renormalization and factorization scale we use $\mu = \mu_F =
m_t$.  Jets are defined by the hadronic $k_T$-algorithm
\cite{Catani:1993hr}: we first remove the charged lepton and the neutrino
from the event, then we precluster all remaining particles and assign
them to the beamjets or to the hard scattering process.  Particles
which are assigned to the hard scattering process are then clustered
into jets.  For the resolution variable of the hadronic
$k_T$-algorithm we use
\begin{eqnarray} \label{eq:33}
y_{ij} & = & 2\; \mbox{min}\;(p_{Ti}^2, p_{Tj}^2)
\left( \cosh(y_i - y_j) - \cos(\phi_i - \phi_j) \right),
\end{eqnarray}
where $p_{Ti}$ is the transverse momentum, $y_i$ the rapidity and
$\phi_i$ the polar angle of particle $i$.  We recombine two particles
using the $E$-scheme.  For the preclustering we use $d_{cut} = (20
\;\mbox{GeV})^2$. The clustering is done with $y_{cut} = 0.9$.  We
have implemented the finite width of the $W$, $Z$-bosons and of the
top quark by using the complex-mass scheme \cite{Denner:1999gp} which
respects full gauge invariance and which therefore gives a consistent 
description of the finite-width effects in tree-level calculations.
Thus, the masses $m_W$, $m_Z$ and $m_t$ in the partial amplitudes are replaced
according to
\begin{eqnarray} \label{eq:34}
m & \rightarrow & \sqrt{m^2 - i \Gamma m}.
\end{eqnarray}
As a consequence the cosine squared of 
the Weinberg angle also becomes a complex number
\begin{eqnarray} \label{eq:35}
\cos^2 \theta_W & = & 1 - \sin^2 \theta_W = \frac{m_W^2-i \Gamma_W m_W}{m_Z^2 -i \Gamma_Z m_Z}.
\end{eqnarray}
We give our results for single-top production only.  Furthermore 
we concentrate most of our studies on the Tevatron. LHC kinematics
reweights the various processes among each other, the amount of which is not
so much our concern in this paper (see \cite{Stelzer:1998ni} e.g.).

Although we do not combine the partonic subprocesses, and rather
examine them individually, we wish to define their final state
in a semi-realistic manner. Specifically, we keep the 
parton apart in phase space by means of a jet algorithm
(this includes in particular, for the W-gluon channel, 
the beam and the $\bar{b}$ jet). However, we do assume
perfect $b$-tagging, and no mistagging. 

The inclusion of single-antitop production will multiply the 
cross section by a factor of two at the Tevatron.  This is not the 
case at the LHC, which is a
proton-proton collider.  The numerical results are for one light
lepton species only, e.g. $\bar{l}=e^+$ and $\nu = \nu_e$. The
inclusion of the muon-channel multiplies every result
by two.\\
\\
For W-gluon fusion we require three jets, two of them $b$-tagged, for
flavor-excitation two jets with one $b$-tag, whereas for
the $s$-channel process we require
two $b$-tagged jets. For simplicity we assume a $b$-tagging efficiency of $100 \%$,
and that we know the longitudinal momentum component of the
neutrino\footnote{E.g. from imposing the $W$ mass constraint on the
neutrino plus lepton invariant mass \cite{Yuan:1990tc}.}.\\
\\
In Table \ref{num_res1} we give the numerical results for the total
cross section with the cuts described above for W-gluon fusion,
flavor-excitation and $s$-channel process at the Tevatron
(first column).  In the second column we required in addition that the
invariant mass of the decay products of the top reconstruct to within
20 GeV to the top quark mass.  The third column contains the results
in the narrow width approximation.  Table \ref{num_res2} shows the
corresponding results for the LHC.
\begin{table}
\begin{center}
\begin{tabular}{|c|c|c|c|} \hline
 Tevatron & $\sigma_{tot}$ & $| m_{\nu\bar{l}b} - m_t | < 20 \;\mbox{GeV}$ & narrow width \\ \hline
$W g$ & $15.0 \pm 0.4 \; \mbox{fb}$ & $14.3 \pm 0.3 \; \mbox{fb}$ & $14.5 \pm 0.1 \; \mbox{fb}$\\ \hline
$W b$ & $87 \pm 1 \; \mbox{fb}$ & $85 \pm 2 \; \mbox{fb}$ & $87 \pm 1 \; \mbox{fb}$\\ \hline
$q \bar{q}$ & $46 \pm 1 \; \mbox{fb}$ & $32.3 \pm 0.3 \; \mbox{fb}$ & $29.0 \pm 0.2 \; \mbox{fb}$\\ \hline
\end{tabular}
\end{center}
\caption{\label{num_res1}
Numerical results for Tevatron at $2 \; \mbox{TeV}$.}
\end{table}
\begin{table}
\begin{center}
\begin{tabular}{|c|c|c|c|} \hline
 LHC & $\sigma_{tot}$ & $| m_{\nu\bar{l}b} - m_t | < 20 \;\mbox{GeV}$ & narrow width \\ \hline
$W g$ & $4.6 \pm 0.2 \; \mbox{pb} $ & $4.5 \pm 0.4 \; \mbox{pb}$ & $4.6 \pm 0.1 \; \mbox{pb} $\\ \hline
$W b$ & $13.1 \pm 0.3 \; \mbox{pb}$ & $13.0 \pm 0.4 \; \mbox{pb}$ & $13.3 \pm 0.1 \; \mbox{pb}$\\ \hline
$q \bar{q}$ & $685 \pm 19 \; \mbox{fb}$ & $479 \pm 16 \; \mbox{fb}$ & $432 \pm 4 \; \mbox{fb}$\\ \hline
\end{tabular}
\end{center}
\caption{\label{num_res2}
Numerical results for the LHC at $14 \; \mbox{TeV}$.}
\end{table}
From Table \ref{num_res1} and Table \ref{num_res2} we see that the
narrow width approximation describes the cross section very well for
W-gluon fusion and flavor excitation.  The approximation is less
satisfactory for s-channel process. Here non-resonant terms seem to
give a more sizeable contribution.
\begin{figure}
\centerline{
\epsfig{file=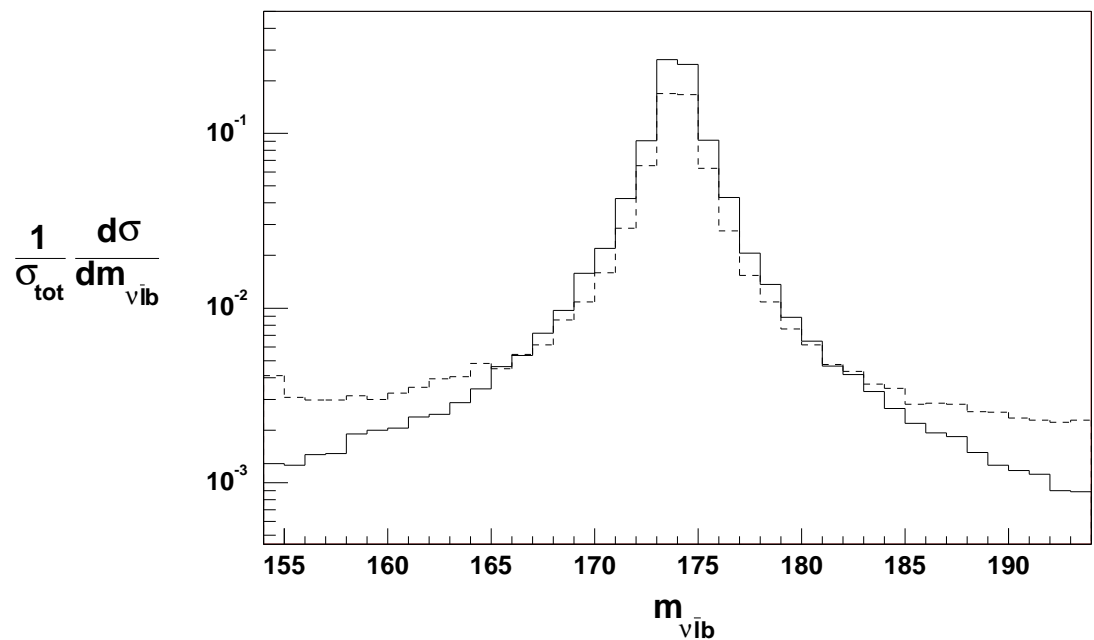, width=16cm} }
\caption{\label{plot1} \it The normalized $m_{\nu\bar{l}b}$ distribution for W-gluon fusion (solid) and the
s-channel process (dashed) at the Tevatron.}
\end{figure}
This can also be seen in Fig.~\ref{plot1},
which shows the distribution in the invariant mass
$m_{\nu\bar{l}b}$ for W-gluon fusion and the s-channel process
at the Tevatron. 
\begin{figure}
\centerline{
\epsfig{file=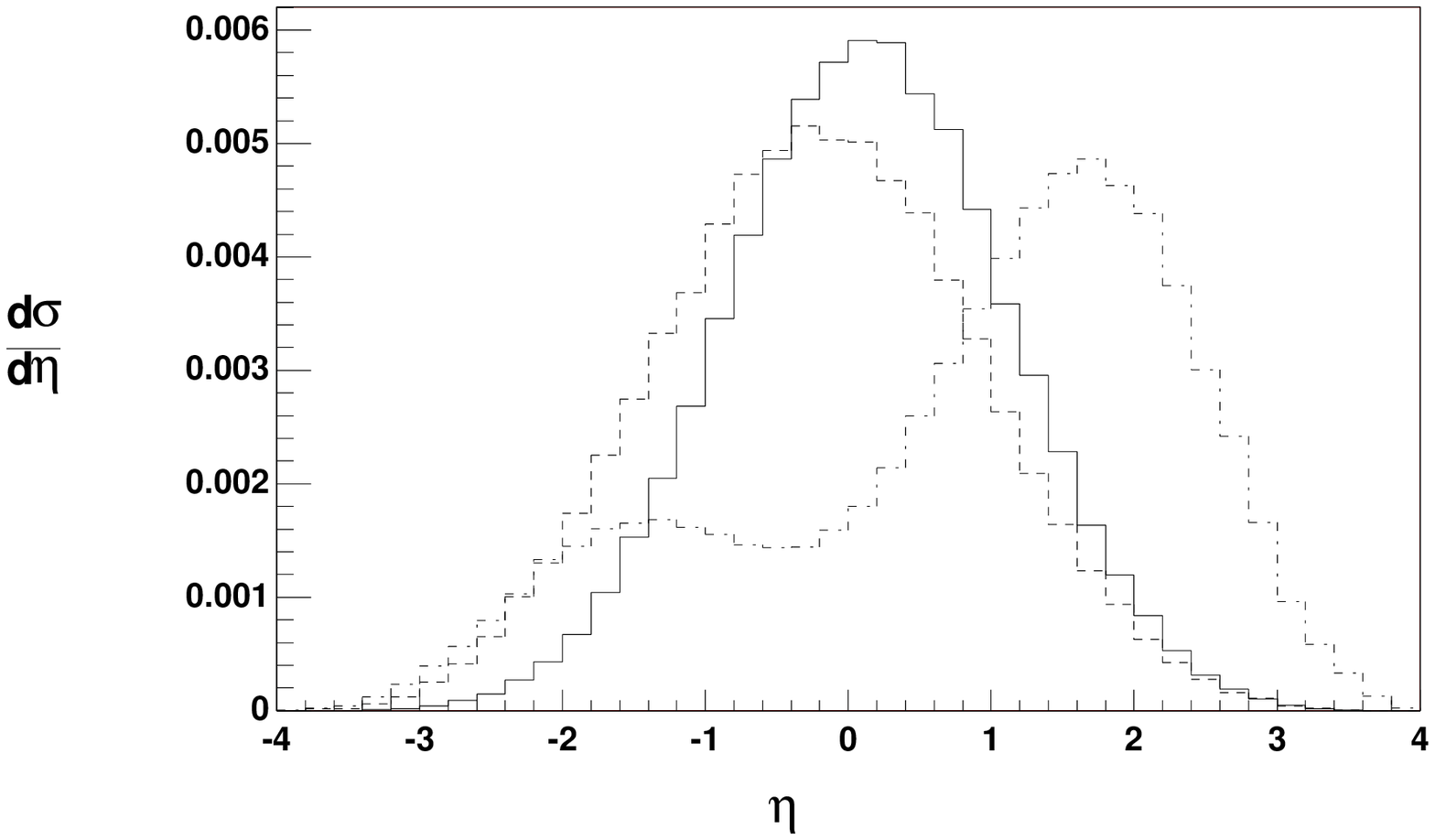, width=16cm}
}
\caption{\label{plot2} \it The pseudo-rapidity distribution for W-gluon fusion 
of the $\bar{b}$-quark (dashed), the $b$-quark (solid) and the light quark $q$ (dot-dashed) 
at the Tevatron.}
\end{figure}
\\
\\
In Fig.~\ref{plot2} we show for the W-gluon fusion process at the Tevatron
the distribution of the pseudorapidities
for the $\bar{b}$-quark, the $b$-quark and the light quark $q$. 
The distribution for the $\bar{b}$-quark is slightly peaked in the backward region,
the $b$-quark is almost central and the light quark goes dominantly in the forward region. 
Note that the jet algorithm suppresses $\bar{b}$'s at sizeable negative pseudo-rapidities.
These distributions essentially agree with Fig.~7 in \cite{Ellis:1992yw} and
Fig.~8 in \cite{Heinson:1997zm}. 
\begin{figure}
\centerline{
\epsfig{file=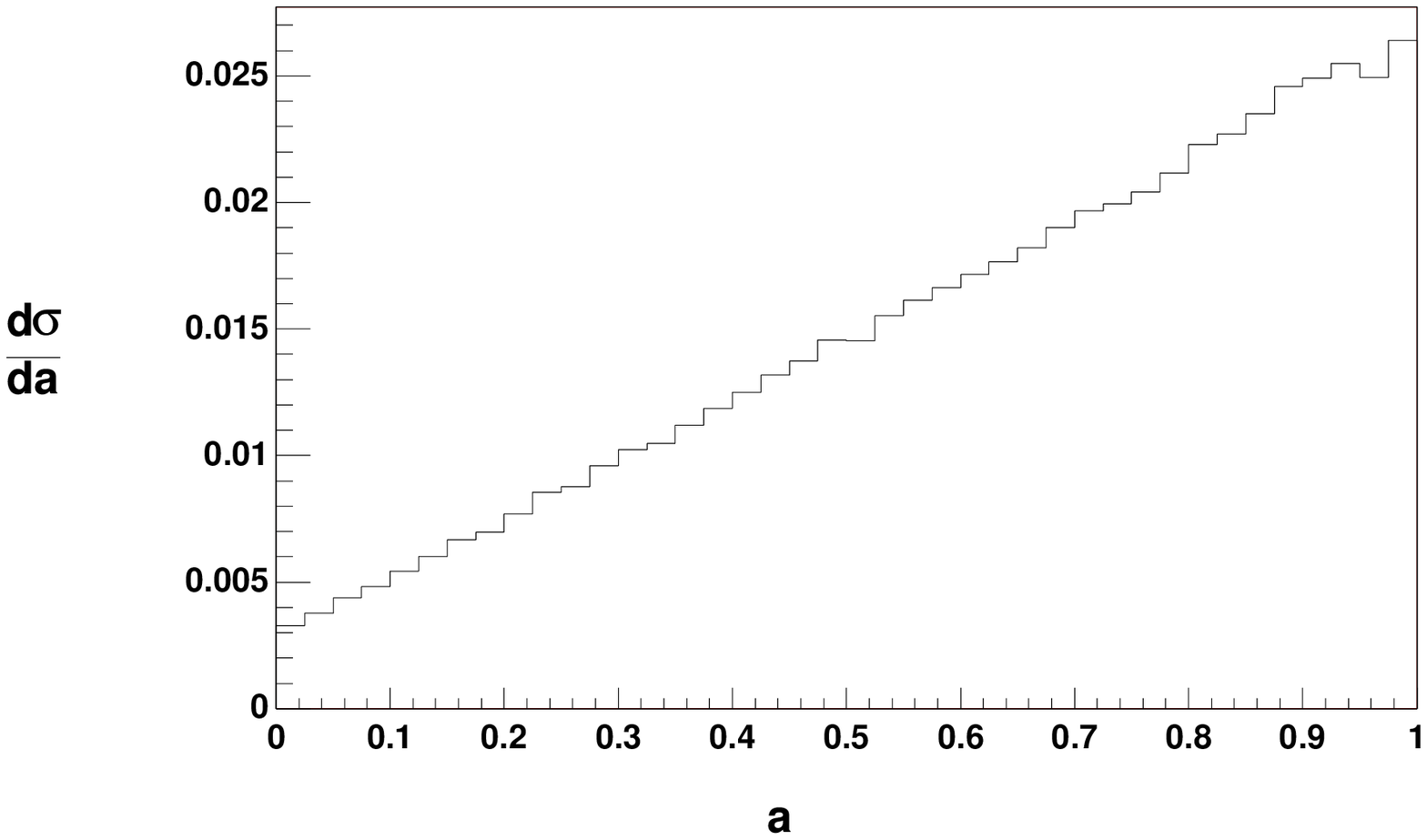, width=16cm} }
\centerline{
\epsfig{file=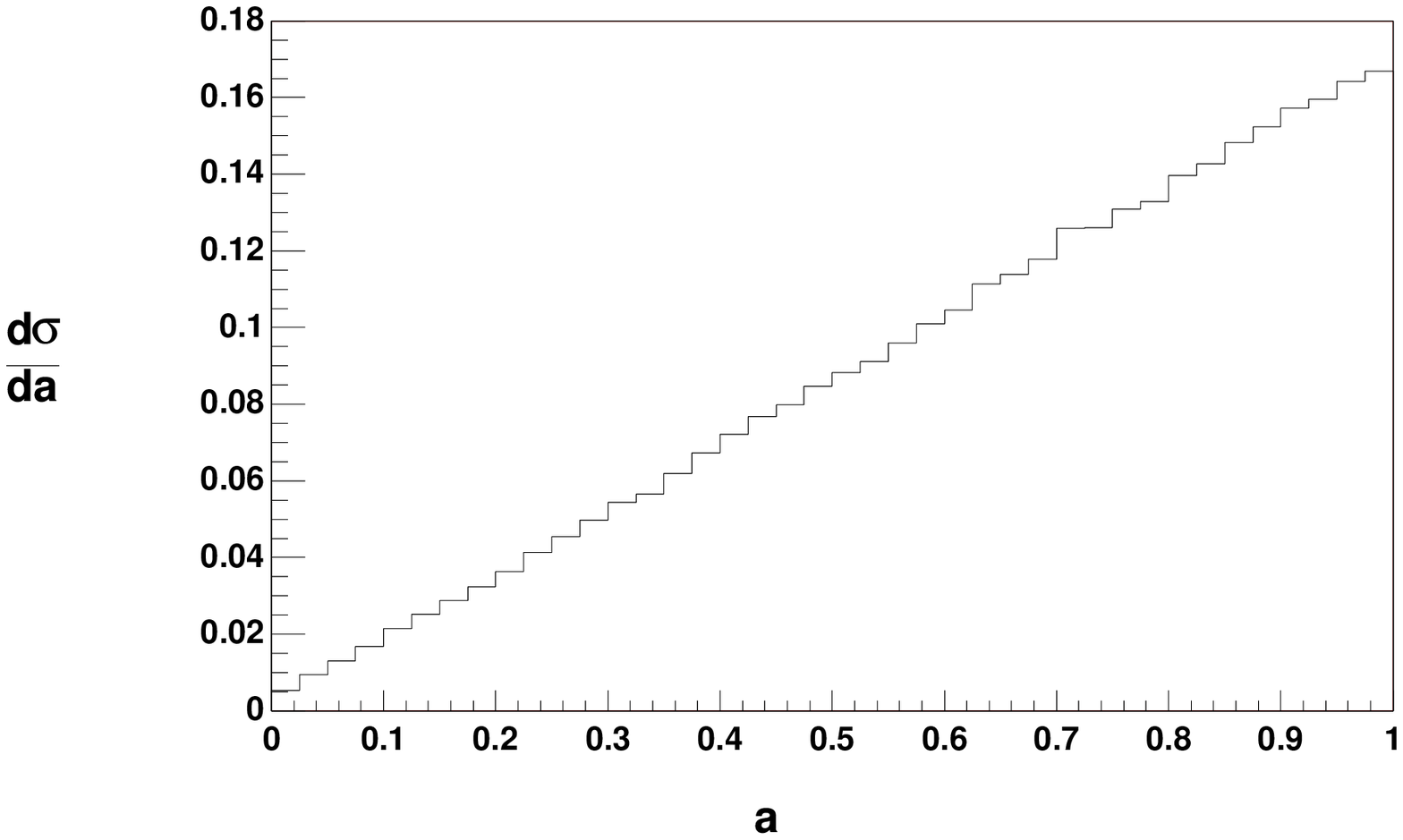, width=16cm}
}
\caption{\label{plot3} \it The distribution for the angular correlation $a$ for W-gluon fusion (top)
and flavor excitation (bottom) at the Tevatron.}
\end{figure}
\begin{figure}
\centerline{
\epsfig{file=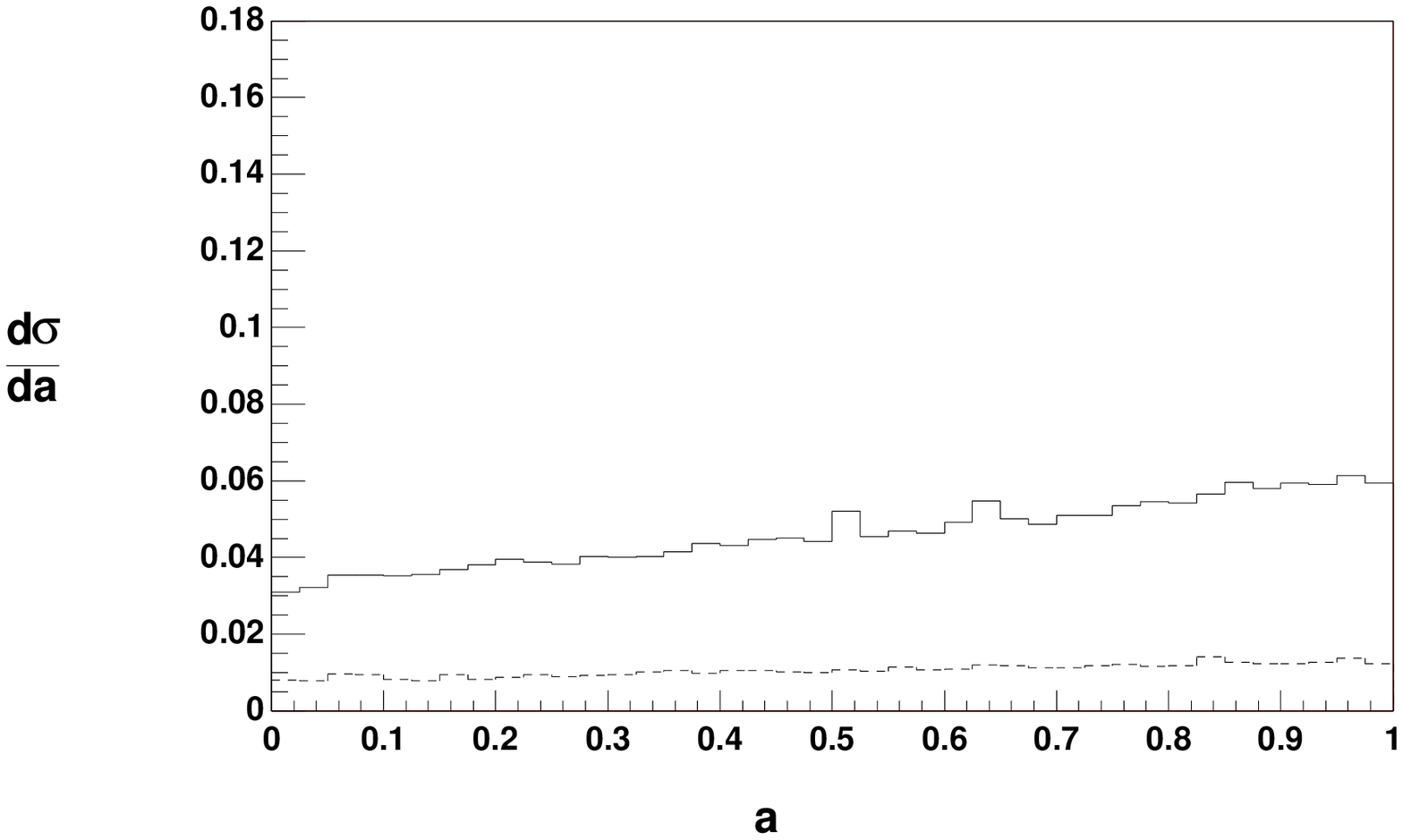, width=16cm}
}
\caption{\label{plot4} \it The distribution for the angular correlation $a$ 
for the ``QCD + Weak'' background to flavor excitation at the Tevatron.
We show the distribution with (dashed) and without (solid) a cut on the invariant mass
$|m_{\nu\bar{l}b}-m_t| < 20\;\mbox{GeV}$.
}
\end{figure}
\\
\\
In W-gluon fusion or flavor excitation the produced top quark is
highly polarized along the direction of the $\bar{d}$-quark
\cite{Mahlon:1997pn,Mahlon:1999gz}.  
Furthermore the cross section at the Tevatron receives
the dominant contribution from the configuration where the $u$-quark
is in the initial state and the $\bar{d}$-quark in the final state,
which in turn produces the non-$b$ tagged jet $q$.  One considers
therefore the variable
\begin{eqnarray} \label{eq:36}
a & = & \frac{1}{2} \left( 1 + \cos \theta_{q\bar{l}} \right)
\end{eqnarray}
where $\theta_{q\bar{l}}$ is the angle between the non-$b$ tagged jet
and the charged lepton in the rest frame of $p_\nu+p_{\bar{l}}+p_b$
\cite{Mahlon:1996zn}.
(If a top is produced, $p_\nu+p_{\bar{l}}+p_b$ corresponds to its
four-momentum.)  
For the angular correlation of a decaying top quark one has \cite{Jezabek:1994qs}
\begin{eqnarray} \label{eq:37}
\frac{d\sigma}{da} & = & \sigma \left( 2 P a + (1-P) \right),
\end{eqnarray}
where $P$ is the polarization of the top quark along the spin axis defined by the spectator
jet $q$. 
For a 100\% polarized top quark one has therefore $d\sigma/da \sim a$. 
Fig.~\ref{plot3} shows that this relation is fullfilled to a very
good approximation for flavour excitation.
For W-gluon fusion we obtain the polarization along the spectator jet axis from the value
at $a=0$:
\begin{eqnarray}
P & = & 1 - \left. \frac{1}{\sigma} \frac{d\sigma}{da} \right|_{a=0}
\end{eqnarray}
With the total cross section from Table 1 we find $P=77 \%$ for
W-gluon fusion.  The background is expected to give a flat
distribution.  For flavor excitation we show in fact the result for
the ``QCD + Weak'' background process in Fig.~\ref{plot4}, discussed
below (\ref{eq:16}), whose $a$ dependence after imposing a cut on
the invariant mass $m_{\nu\bar{l}b}$ is flat, as expected.  Our result
is similar to that shown in Fig.~5 of \cite{Stelzer:1998ni},
which shows the same linear correlation of the signal (which is somewhat
different from ours by employing a vetoed $\bar{b}$ jet), and shows
the (flat) $a$ dependence of their more extensively treated background
as well.

We suggest that this clear correlation may provide an alternative and
attractive way to infer the visible W-gluon fusion or flavor excitation
cross section (defined here through the criteria given above)
for single-top production from the slope of the distribution.  
The slope is given by
\begin{eqnarray}
2 P_{signal} \sigma_{signal} + 2 P_{background} \sigma_{background}
\end{eqnarray}
Assuming that $P_{background} \sigma_{background}$ is small and that
$P_{signal}$ may reliable be estimated from theory the visible cross
section for the signal can be inferred from the slope by measuring two
or more points of the $a$ distribution and extrapolating the
distribution to a straight line.  Although in principle of course any
distribution may serve to infer the corresponding inclusive cross
section, the $a$ distribution seems particularly attractive due to its
simple shape.

\section{Conclusions}

We have presented the complete $O(g_w^4)$ and $O(g g_w^4)$
helicity amplitudes for processes
whose final state results from the hadroproduction and
semileptonic decay of a single top.

As only a small subset
of graphs actually contain a top quark line in each process,
we have examined, for three of these processes, to what extent the
top quark dominates, and verified that for each
process the top quark presence is manifested by a clear peak
in the $m_{\nu\bar{l}b}$ distribution.
We have studied various kinematic distribution of 
final state particles, and verified the correlation of 
the lepton angular distribution with the top quark
polarization \cite{Mahlon:1997pn,Stelzer:1998ni,Mahlon:1999gz}. 
The actual identification of a single top signal is
a matter of careful definition, requiring full use of
the kinematic and flavor characteristics of the final
state, and a proper determination
of the acceptance and background \cite{Stelzer:1998ni,Belyaev:1998dn,Liu:1999ra}. 
In an idealized 
analyses, we have thus verified that the sensitivity
of the full amplitudes to top quark mass, charged-current
coupling strength and handedness are preserved in these amplitudes, even
though most diagrams that contribute to them do not 
contain a top quark. 

\subsection*{Acknowledgments}

We would like to thank Brian Harris and Zack Sullivan for 
illuminating discussions. This work is part of the research program of the
Foundation for Fundamental Research of Matter (FOM) and
the National Organization for Scientific Research (NWO).

\begin{appendix}

\section{Helicity amplitudes for W-gluon fusion}

We give here the helicity amplitudes for the configurations
$(p_3^-,p_4^+,p_5^+)$, $(p_3^-,p_4^+,p_5^-)$,\\
 $(p_3^+,p_4^-,p_5^+)$ and $(p_3^+,p_4^-,p_5^-)$.
\begin{eqnarray} \label{eq:38}
A^{(1,1)}_{Wg} & = & \frac{4 \sqrt{2} i B^{(1,1)}_{Wg}}{(s_{12}-m_W^2)(s_{67}-m_W^2)},
\end{eqnarray}
\begin{eqnarray} \label{eq:39}
B^{(1,1)}_{Wg}(p_3^-,p_4^+,p_5^+) & = & \frac{1}{s_{123}-m^2} \frac{\l 31 \r}{\l 35 \r}
\left( \frac{\l 3- | 4+5 | 7- \r \l 6- | 1+3 | 2- \r}{\l 54 \r} \right. \nonumber \\
& & \left. + \frac{[ 47 ]}{s_{467}-m^2} \left( [ 21 ] \l 13 \r \l 6- | 4+7 | 5- \r - m^2 [ 25 ] \l 36 \r \right) \right),     \nonumber \\
B^{(1,1)}_{Wg}(p_3^-,p_4^+,p_5^-) & = & \frac{1}{s_{467}-m^2} \frac{[ 74 ]}{[ 54 ]}
\left( \frac{\l 1- | 3+5 | 4- \r \l 6- | 4+7 | 2- \r}{[ 35 ]} \right. \nonumber \\
& & \left. + \frac{\l 13 \r}{s_{123} - m^2} \left( \l 67 \r [ 74 ] \l 5- | 1+3 | 2- \r - m^2 \l 56 \r [ 24 ] \right)
\right),         \nonumber \\
B^{(1,1)}_{Wg}(p_3^+,p_4^-,p_5^+) & = & 0,        \nonumber \\
B^{(1,1)}_{Wg}(p_3^+,p_4^-,p_5^-) & = & 0, \nonumber
\end{eqnarray}
\begin{eqnarray} \label{eq:40}
A^{(1,2)}_{Wg} & = & \frac{4 \sqrt{2} i B^{(1,2)}_{Wg}}{(s_{12}-m_W^2) s_{345} s_{127}},
\end{eqnarray}
\begin{eqnarray} \label{eq:41}
B^{(1,2)}_{Wg}(p_3^-,p_4^+,p_5^+) & = & \frac{\l 36 \r}{\l 54 \r \l 35 \r} [ 27 ] \l 1- | (2+7) (4+5) | 3+ \r,      \nonumber \\
B^{(1,2)}_{Wg}(p_3^-,p_4^+,p_5^-) & = & \frac{[ 27 ]}{[ 54 ] [ 53 ] } 
   \l 6- | 3+5 | 4- \r \l 1- | 2+7 | 4- \r,  \nonumber \\
B^{(1,2)}_{Wg}(p_3^+,p_4^-,p_5^+) & = & \frac{ \l 64 \r}{\l 45 \r \l 35 \r} 
   [ 27 ] \l 4- | (3+5) (2+7) | 1+ \r,         \nonumber \\
B^{(1,2)}_{Wg}(p_3^+,p_4^-,p_5^-) & = &  \frac{[ 27 ]}{[ 45 ] [ 53 ]} \l 1- | 2+7 | 3- \r \l 6- | 4+5 | 3- \r, \nonumber
\end{eqnarray}
\begin{eqnarray} \label{eq:42}
A^{(1,3)}_{Wg} & = & \frac{4 \sqrt{2} i B^{(1,3)}_{Wg}}{(s_{12}-m_W^2) s_{345} s_{126} },
\end{eqnarray}
\begin{eqnarray} \label{eq:43}
B^{(1,3)}_{Wg}(p_3^-,p_4^+,p_5^+) & = & \frac{\l 61 \r}{\l 45 \r \l 35 \r} \l 3- | 1+6 | 2- \r \l 3- | 4+5 | 7- \r,      \nonumber \\
B^{(1,3)}_{Wg}(p_3^-,p_4^+,p_5^-) & = & \frac{[ 47 ]}{[ 54 ] [ 53 ]}
   \l 61 \r \l 4+ | (3+5) (1+6) | 2- \r,       \nonumber \\
B^{(1,3)}_{Wg}(p_3^+,p_4^-,p_5^+) & = & \frac{\l 61 \r}{\l 45 \r \l 53 \r}
   \l 4- | 1+6 | 2- \r \l 4- | 3+5 | 7- \r,       \nonumber \\
B^{(1,3)}_{Wg}(p_3^+,p_4^-,p_5^-) & = &  \frac{[ 37 ]}{[ 54 ] [ 53 ]} \l 61 \r \l 2+ | (1+6) (4+5) | 3- \r, \nonumber 
\end{eqnarray}
\begin{eqnarray} \label{eq:44}
A^{(1,4)}_{Wg} & = & \frac{4 \sqrt{2} i B^{(1,4)}_{Wg}}{(s_{12}-m_W^2) (s_{67}-m_W^2) s_{345} },
\end{eqnarray}
\begin{eqnarray} \label{eq:45}
\lefteqn{B^{(1,4)}_{Wg}(p_3^-,p_4^+,p_5^+) = \frac{1}{\l 54 \r \l 35 \r} \left(
   \l 31 \r  \l 6- | 1+2 | 7- \r \l 3- | 4+5 | 2 - \r \right. } & & \nonumber \\
  & & \left. - \l 36 \r  \l 1- | 6+7 | 2- \r \l 3- | 4+5 | 7- \r
  + \l 16 \r [ 72 ] \l 3- | (6+7) (4+5) | 3+ \r \right),       \nonumber \\
\lefteqn{B^{(1,4)}_{Wg}(p_3^-,p_4^+,p_5^-) = \frac{1}{[ 53 ] [ 54 ]} \left(
   [ 24 ] \l 6- | 1+2 | 7- \r \l 1- | 3+5 | 4- \r \right. } & & \nonumber \\
  & & \left.   - [ 74 ] \l 1- | 6+7 | 2- \r \l 6- | 3+5 | 4- \r
  - [ 72 ] \l 16 \r \l 4+ | (6+7) (3+5) | 4- \r \right),    \nonumber \\
\lefteqn{B^{(1,4)}_{Wg}(p_3^+,p_4^-,p_5^+) = \frac{1}{\l 45 \r \l 53 \r} \left(
   \l 14 \r \l 6- | 1+2 | 7- \r \l 4- | 3+5 | 2- \r \right. } & & \nonumber \\
  & & \left. - \l 64 \r \l 1- | 6+7 | 2- \r \l 4- | 3+5 | 7- \r
  + \l 16 \r [ 72 ] \l 4- | (3+5) (6+7) | 4+ \r \right),        \nonumber \\
\lefteqn{B^{(1,4)}_{Wg}(p_3^+,p_4^-,p_5^-) = \frac{1}{[ 54 ] [53 ]} \left(
   [ 32 ] \l 6- | 1+2 | 7- \r \l 1- | 4+5 | 3 - \r \right. } & & \nonumber \\
  & & \left. - [ 37 ] \l 1- | 6+7 | 2- \r \l 6- | 4+5 | 3- \r
  + \l 16 \r [ 72 ] \l 3+ | (6+7) (4+5) | 3- \r \right), \nonumber
\end{eqnarray}
\begin{eqnarray} \label{eq:46}
A^{(1,5)}_{Wg} & = & \frac{4 \sqrt{2} i B^{(1,5)}_{Wg}}{(s_{67}-m_W^2) s_{345} s_{167}},
\end{eqnarray}
\begin{eqnarray} \label{eq:47}
B^{(1,5)}_{Wg}(p_3^-,p_4^+,p_5^+) & = & \frac{\l 16 \r}{\l 35 \r \l 45 \r} 
 \l 3- | 1+6 | 7- \r \l 3- | 4+5 | 2- \r, \nonumber \\
B^{(1,5)}_{Wg}(p_3^-,p_4^+,p_5^-) & = & \frac{[ 24 ]}{[ 45 ] [ 35 ]} 
 \l 16 \r \l 7+ | (1+6) (3+5) |4- \r, \nonumber \\ 
B^{(1,5)}_{Wg}(p_3^+,p_4^-,p_5^+) & = & - \frac{ \l 16 \r}{\l 35 \r \l 45 \r}
 \l 4- | 1+6 | 7- \r \l 4- | 3+5 | 2- \r, \nonumber \\
B^{(1,5)}_{Wg}(p_3^+,p_4^-,p_5^-) & = & - \frac{[ 23 ]}{[ 35 ] [ 45 ]} 
 \l 16 \r \l 7+ | (1+6) (4+5) | 3- \r, \nonumber 
\end{eqnarray}
\begin{eqnarray} \label{eq:48}
A^{(1,6)}_{Wg} & = & \frac{4 \sqrt{2} i B^{(1,6)}_{Wg}}{(s_{67}-m_W^2) s_{345} s_{267}},
\end{eqnarray}
\begin{eqnarray} \label{eq:49}
B^{(1,6)}_{Wg}(p_3^-,p_4^+,p_5^+) & = & \frac{\l 13 \r}{\l 35 \r \l 45 \r}
 [ 27 ] \l 3- | (4+5) (2+7) | 6+ \r, \nonumber \\
B^{(1,6)}_{Wg}(p_3^-,p_4^+,p_5^-) & = & - \frac{[ 27 ]}{[ 35 ] [ 45 ]}
 \l 1- | 3+5 | 4- \r \l 6- | 2+7 | 4- \r, \nonumber \\
B^{(1,6)}_{Wg}(p_3^+,p_4^-,p_5^+) & = & - \frac{\l 14 \r}{\l 35 \r \l 45 \r}
 [ 27 ] \l 4- | (3+5) (2+7) | 6+ \r, \nonumber \\
B^{(1,6)}_{Wg}(p_3^+,p_4^-,p_5^-) & = & \frac{[ 27 ]}{[ 35 ] [ 45 ]}
 \l 1- | 4+5 | 3- \r \l 6- | 2+7 | 3- \r, \nonumber 
\end{eqnarray}
\begin{eqnarray} \label{eq:50}
A^{(2,1)}_{Wg} & = & \frac{4 \sqrt{2} i B^{(2,1)}_{Wg}}{(s_{12}-m_W^2)(s_{567}-m_W^2)(s_{123}-m^2)},
\end{eqnarray}
\begin{eqnarray} \label{eq:51}
B^{(2,1)}_{Wg}(p_3^-,p_4^+,p_5^+) & = & \frac{\l 31 \r}{\l 65 \r \l 75 \r} 
   \l 6- | 1+3 | 2- \r \l 6- | 5+7 | 4- \r,      \nonumber \\
B^{(2,1)}_{Wg}(p_3^-,p_4^+,p_5^-) & = & \frac{[ 47 ]}{[ 57 ] [ 56 ]} 
   \l 31 \r \l 2+ | (1+3) (5+6) | 7- \r,       \nonumber \\
B^{(2,1)}_{Wg}(p_3^+,p_4^-,p_5^+) & = & 0,        \nonumber \\
B^{(2,1)}_{Wg}(p_3^+,p_4^-,p_5^-) & = & 0, \nonumber
\end{eqnarray}
\begin{eqnarray} \label{eq:52}
A^{(2,2)}_{Wg} & = & \frac{4 \sqrt{2} i B^{(2,2)}_{Wg}}{(s_{12}-m_W^2) s_{34}},
\end{eqnarray}
\begin{eqnarray} \label{eq:53}
\lefteqn{B^{(2,2)}_{Wg}(p_3^-,p_4^+,p_5^+) = } & & \nonumber \\
& & \frac{1}{s_{346}} \frac{\l 63 \r}{\l 65 \r} \left(
    \frac{\l 1- | 3+6 | 4- \r \l 6- | 5+7 | 2- \r}{\l 57 \r} 
   - \frac{[ 43 ] \l 36 \r}{s_{127}} [ 27 ] \l 1- | 2+7 | 5- \r \right),       \nonumber \\
\lefteqn{B^{(2,2)}_{Wg}(p_3^-,p_4^+,p_5^-) = } & & \nonumber \\
& & -\frac{1}{s_{127}} \frac{[ 27 ]}{[ 57 ]} \left(
   \frac{\l 3- | 5+6 | 7- \r \l 1- | 2+7 | 4- \r}{[ 56 ]} 
   + \frac{\l 63 \r \l 12 \r [ 27 ] \l 5- | 3+6 | 4- \r}{s_{346}}  \right),       \nonumber \\
\lefteqn{B^{(2,2)}_{Wg}(p_3^+,p_4^-,p_5^+) = } & & \nonumber \\
& & \frac{1}{s_{346}} \frac{\l 64 \r}{\l 65 \r} \left(
    \frac{\l 1- | 4+6 | 3- \r \l 6- | 5+7 | 2- \r}{\l 57 \r}
   - \frac{[ 34 ] \l 46 \r}{s_{127}} [ 27 ] \l 1- | 2+7 | 5- \r \right),         \nonumber \\
\lefteqn{B^{(2,2)}_{Wg}(p_3^+,p_4^-,p_5^-) = } & & \nonumber \\
& & -\frac{1}{s_{127}} \frac{[ 27 ]}{[ 57 ]} \left(
   \frac{\l 4- | 5+6 | 7- \r \l 1- | 2+7 | 3- \r}{[ 56 ]} 
   + \frac{\l 64 \r \l 12 \r [ 27 ] \l 5- | 4+6 | 3- \r}{s_{346}}  \right), \nonumber 
\end{eqnarray}
\begin{eqnarray} \label{eq:54}
A^{(2,3)}_{Wg} & = & \frac{4 \sqrt{2} i B^{(2,3)}_{Wg}}{(s_{12}-m_W^2) s_{34}},
\end{eqnarray}
\begin{eqnarray} \label{eq:55}
\lefteqn{B^{(2,3)}_{Wg}(p_3^-,p_4^+,p_5^+) = } & & \nonumber \\
& & \frac{1}{s_{126}} \frac{\l 61 \r}{\l 65 \r} \left(
   \frac{ \l3- | 1+6 | 2- \r \l 6- | 5+7 | 4- \r}{\l 57 \r}
   - \frac{[ 21 ] \l 16 \r [ 47 ] \l 3- | 4+7 | 5- \r}{s_{347}} \right),        \nonumber \\
\lefteqn{B^{(2,3)}_{Wg}(p_3^-,p_4^+,p_5^-) = } & & \nonumber \\
& & -\frac{1}{s_{347}} \frac{[ 47 ]}{[ 57 ]} \left(
   \frac{\l 1- | 5+6 | 7-  \r \l 3- | 4+7 | 2- \r}{[ 56 ]}
   - \frac{\l 61 \r [ 74 ] \l 34 \r \l 5- | 1+6 | 2- \r}{s_{126}} \right),       \nonumber \\
\lefteqn{B^{(2,3)}_{Wg}(p_3^+,p_4^-,p_5^+) = } & & \nonumber \\
& & \frac{1}{s_{126}} \frac{\l 61 \r}{\l 65 \r} \left(
   \frac{\l 4- | 1+6 | 2- \r \l 6- | 5+7 | 3- \r}{\l 57 \r}
   - \frac{[ 21 ] \l 16 \r [ 37 ] \l 4- | 3+7 | 5- \r}{s_{347}} \right),        \nonumber \\
\lefteqn{B^{(2,3)}_{Wg}(p_3^+,p_4^-,p_5^-) = } & & \nonumber \\
& & -\frac{1}{s_{347}} \frac{[ 37 ]}{[ 57 ]} \left(
   \frac{\l 1- | 5+6 | 7- \r \l 4- | 3+7 | 2- \r}{[ 56 ]}
   - \frac{\l 61 \r [ 73 ] \l 43 \r \l 5- | 1+6 | 2- \r}{s_{126}} \right), \nonumber
\end{eqnarray}
\begin{eqnarray} \label{eq:56}
A^{(2,4)}_{Wg} & = & \frac{4 \sqrt{2} i B^{(2,4)}_{Wg}}{(s_{12}-m_W^2)(s_{567}-m_W^2) s_{34}},
\end{eqnarray}
\begin{eqnarray} \label{eq:57}
\lefteqn{B^{(2,4)}_{Wg}(p_3^-,p_4^+,p_5^+) = -\frac{1}{\l 75 \r \l 65 \r} \left(
   \l 63 \r \l 1- | 3+4 | 2- \r \l 6- | 5+7 | 4- \r \right. } & & \nonumber \\
   & & \left. - \l 61 \r \l 3- | 1+2 | 4- \r \l 6- | 5+7 | 2- \r
   + \l 13 \r [ 42 ] \l 6- | (1+2) (5+7) | 6+ \r \right),       \nonumber \\
\lefteqn{B^{(2,4)}_{Wg}(p_3^-,p_4^+,p_5^-) = -\frac{1}{[ 57 ] [ 65 ]} \left(
   [ 47 ] \l 1- | 3+4 | 2- \r \l 3- | 5+6 | 7- \r \right. } & & \nonumber \\
   & & \left. - [ 27 ] \l 3- | 1+2 | 4- \r \l 1- | 5+6 | 7- \r
   + [ 42 ] \l 13 \r \l 7+ | (5+6) (1+2) | 7 - \r \right),       \nonumber \\
\lefteqn{B^{(2,4)}_{Wg}(p_3^+,p_4^-,p_5^+) = -\frac{1}{\l 75 \r \l 65 \r} \left(
   \l 64 \r \l 1- | 3+4 | 2- \r \l 6- | 5+7 | 3- \r \right. } & & \nonumber \\
   & & \left. - \l 61 \r \l 4- | 1+2 | 3- \r \l 6- | 5+7 | 2- \r
   + \l 14 \r [ 32 ] \l 6- | (1+2) (5+7) | 6+ \r \right),         \nonumber \\
\lefteqn{B^{(2,4)}_{Wg}(p_3^+,p_4^-,p_5^-) = -\frac{1}{[ 57 ] [ 65 ]} \left(
   [ 37 ] \l 1- | 3+4 | 2- \r \l 4- | 5+6 | 7- \r \right. } & & \nonumber \\
   & & \left. - [ 27 ] \l 4- | 1+2 | 3- \r \l 1- | 5+6 | 7- \r
   + [ 32 ] \l 1 4 \r \l 7+ | (5+6) (1+2) | 7- \r \right), \nonumber 
\end{eqnarray}
\begin{eqnarray} \label{eq:58}
A^{(2,5)}_{Wg} & = & \frac{4 \sqrt{2} i B^{(2,5)}_{Wg}}{(s_{567}-m_W^2) s_{34} s_{234}},
\end{eqnarray}
\begin{eqnarray} \label{eq:59}
B^{(2,5)}_{Wg}(p_3^-,p_4^+,p_5^+) & = &  \frac{\l 16 \r}{\l 56 \r \l 57 \r}
 [ 24 ] \l 6- | (5+7) (2+4) | 3+ \r, \nonumber \\
B^{(2,5)}_{Wg}(p_3^-,p_4^+,p_5^-) & = & - \frac{[ 24 ]}{[ 56 ] [ 57 ]}
 \l 1- | 5+6 | 7- \r \l 3- | 2+4 | 7- \r, \nonumber \\
B^{(2,5)}_{Wg}(p_3^+,p_4^-,p_5^+) & = & \frac{\l 16 \r}{\l 56 \r \l 57 \r}
 [ 23 ] \l 6- | (5+7) (2+3) | 4+ \r, \nonumber \\
B^{(2,5)}_{Wg}(p_3^+,p_4^-,p_5^-) & = & - \frac{[ 23 ]}{[ 56 ] [ 57 ]}
 \l 1- | 5+6 | 7- \r \l 4- | 2+3 | 7- \r, \nonumber  
\end{eqnarray}
\begin{eqnarray} \label{eq:60}
A^{(2,6)}_{Wg} & = & \frac{4 \sqrt{2} i B^{(2,6)}_{Wg}}{(s_{567}-m_W^2) s_{34} s_{134}},
\end{eqnarray}
\begin{eqnarray} \label{eq:61}
B^{(2,6)}_{Wg}(p_3^-,p_4^+,p_5^+) & = & \frac{\l 13 \r}{\l 56 \r \l 57 \r}
 \l 6- | 1+3 | 4- \r \l 6- | 5+7 | 2- \r, \nonumber \\
B^{(2,6)}_{Wg}(p_3^-,p_4^+,p_5^-) & = & \frac{[ 27 ]}{[ 56 ] [ 57 ]}
 \l 13 \r \l 4+ | (1+3) ( 5+6) | 7- \r, \nonumber \\
B^{(2,6)}_{Wg}(p_3^+,p_4^-,p_5^+) & = & \frac{\l 14 \r}{\l 56 \r \l 57 \r}
 \l 6- | 1+4 | 3- \r \l 6- | 5+7 | 2- \r, \nonumber \\
B^{(2,6)}_{Wg}(p_3^+,p_4^-,p_5^-) & = & \frac{[ 27 ]}{[ 56 ] [ 57 ]}
 \l 14 \r \l 3+ | (1+4) (5+6) | 7- \r. \nonumber
\end{eqnarray}

\section{Helicity amplitudes for flavor excitation and the $s$-channel}

We give here the helicity amplitudes for the configurations
$(p_3^-,p_4^+)$, and $(p_3^+,p_4^-)$.
\begin{eqnarray} \label{eq:62}
A^{(1,1)}_{Wb} & = & \frac{- 4 i B^{(1,1)}_{Wb}}{(s_{12}-m_W^2)(s_{67}-m_W^2)(s_{123}-m^2)},
\end{eqnarray}
\begin{eqnarray} \label{eq:63}
B^{(1,1)}_{Wb}(p_3^-,p_4^+) & = & \l 31 \r [ 74 ] \l 6- | 1+3 | 2- \r, \nonumber \\
B^{(1,1)}_{Wb}(p_3^+,p_4^-) & = & 0, \nonumber
\end{eqnarray}
\begin{eqnarray} \label{eq:64}
A^{(1,2)}_{Wb} & = & \frac{- 4 i B^{(1,2)}_{Wb}}{(s_{12}-m_W^2) s_{34} s_{346}},
\end{eqnarray}
\begin{eqnarray} \label{eq:65}
B^{(1,2)}_{Wb}(p_3^-,p_4^+) & = & \l 63 \r [ 27 ] \l 1- | 3+6 | 4- \r, \nonumber \\
B^{(1,2)}_{Wb}(p_3^+,p_4^-) & = & \l 64 \r [ 27 ] \l 1- | 4+6 | 3- \r, \nonumber
\end{eqnarray}
\begin{eqnarray} \label{eq:66}
A^{(1,3)}_{Wb} & = & \frac{- 4 i B^{(1,3)}_{Wb}}{(s_{12}-m_W^2) s_{34} s_{126}},
\end{eqnarray}
\begin{eqnarray} \label{eq:67}
B^{(1,3)}_{Wb}(p_3^-,p_4^+) & = & \l 61 \r [ 47 ] \l 3- | 1+6 | 2- \r, \nonumber \\
B^{(1,3)}_{Wb}(p_3^+,p_4^-) & = & \l 61 \r [ 37 ] \l 4- | 1+6 | 2- \r, \nonumber
\end{eqnarray}
\begin{eqnarray} \label{eq:68}
A^{(1,4)}_{Wb} & = & \frac{- 4 i B^{(1,4)}_{Wb}}{(s_{12}-m_W^2)(s_{67}-m_W^2) s_{34}},
\end{eqnarray}
\begin{eqnarray} \label{eq:69}
B^{(1,4)}_{Wb}(p_3^-,p_4^+) & = & - \l 63 \r [ 47 ] \l 1- | 6+7 | 2- \r
                           - \l 13 \r [ 42 ] \l 6- | 3+4 | 7- \r
                           \nonumber \\
& &                        - \l 61 \r [ 27 ] \l 3- | 1+2 | 4- \r, \nonumber \\
B^{(1,4)}_{Wb}(p_3^+,p_4^-) & = & - \l 64 \r [ 37 ] \l 1- | 6+7 | 2- \r
                           - \l 14 \r [ 32 ] \l 6- | 3+4 | 7- \r
                           \nonumber \\
& &                           - \l 61 \r [ 27 ] \l 4- | 1+2 | 3- \r, \nonumber
\end{eqnarray}
\begin{eqnarray} \label{eq:70}
A^{(1,5)}_{Wb} & = & \frac{- 4 i B^{(1,5)}_{Wb}}{(s_{67}-m_W^2) s_{34} s_{167}},
\end{eqnarray}
\begin{eqnarray} \label{eq:71}
B^{(1,5)}_{Wb}(p_3^-,p_4^+) & = & \l 16 \r [ 42 ] \l 3- | 1+6 | 7- \r, \nonumber \\
B^{(1,5)}_{Wb}(p_3^+,p_4^-) & = & \l 16 \r [ 32 ] \l 4- | 1+6 | 7- \r, \nonumber
\end{eqnarray}
\begin{eqnarray} \label{eq:72}
A^{(1,6)}_{Wb} & = & \frac{- 4 i B^{(1,6)}_{Wb}}{(s_{67}-m_W^2) s_{34} s_{134}},
\end{eqnarray}
\begin{eqnarray} \label{eq:73}
B^{(1,6)}_{Wb}(p_3^-,p_4^+) & = & \l 13 \r [ 72 ] \l 6- | 1+3 | 4- \r, \nonumber \\
B^{(1,6)}_{Wb}(p_3^+,p_4^-) & = & \l 14 \r [ 72 ] \l 6- | 1+4 | 3- \r. \nonumber
\end{eqnarray}

\end{appendix}

\end{document}